\documentclass[preprint,showpacs,preprintnumbers,amsmath,amssymb]{revtex4}

\usepackage{graphicx}
\usepackage{dcolumn}
\usepackage{bm}

\graphicspath{{Fig/}}

\begin{document}

\preprint{Manuscript}

\title{Contributions of Al and Ni segregation to the interfacial cohesion of Cu-rich precipitates in ferritic steels}

\author{Yao-Ping Xie}
\author{Shi-Jin Zhao}%
 \email{shijin.zhao@shu.edu.cn}
\affiliation{%
 Institute of Materials Science, School of Materials Science and Engineering, Shanghai University, Shanghai, 200444, China
}%

\date{\today}

\begin{abstract}
We characterise the influence of the segregation behaviours of two
typical alloying elements, aluminium and nickel, on the interfacial
cohesive properties of copper-rich precipitates in ferritic steels,
with a view towards understanding steel embrittlement. The
first-principles method is used to compute the energetic and bonding
properties of aluminium and nickel at the interfaces of the
precipitates and corresponding fracture surfaces. Our results show
the segregation of aluminium and nickel at interfaces of
precipitates are both energetically favourable. We find that the
interfacial cohesion of copper precipitates is enhanced by aluminium
segregation but reduced by nickel segregation. Opposite roles can be
attributed to the different symmetrical features of the valence
states for aluminium and nickel. The nickel-induced interfacial
embrittlement of copper-rich precipitates increase the
ductile-brittle transition temperature (DBTT) of ferritic steels and
provides an explanation of many experimental phenomena, such as the
fact that the shifts of DBTT of reactor pressure vessel steels
depend the copper and nickel content.

\end{abstract}

\pacs{81.40.Np, 68.35.Dv, 64.75.Jk, 81.40.Cd}
\maketitle


\section{\label{sec:level1}Introduction}

It has long been known that the copper content in steels leads to
precipitation hardening. Copper is an element commonly occurring in
steels either as an intentionally added alloying species or as an
impurity. Nanoscale copper-rich precipitates are utilised to provide
substantial precipitation hardening for high-strength low-alloy
steels, which possess excellent impact toughness, corrosion
resistance, and welding properties
\cite{steel-high1,steel-high2,steel-high3}. In contrast, copper-rich
precipitates induce hardening and embrittlement effects in reactor
pressure vessel steels (RPV) after neutron irradiation
\cite{miller2,rpv2,rpv3}, thereby limiting the operational life of
nuclear power plants. Therefore, understanding the properties of
copper-rich precipitates is desirable.

Many investigations have provided insight into the hardening
mechanism that results from copper-rich precipitation in ferritic
steels. Molecular dynamics simulation has suggested that the major
source of precipitation hardening is the dislocation
core-precipitate interaction \cite{ther2,ther3,conment}. Dislocation
core-precipitate interactions tend to induce the loss of screw
dislocation slip systems and the transformation of the copper phase
for larger body center cubic (BCC) copper-rich precipitates (d $>$
3.3 nm) while inducing polarised-to-nonpolarised transitions of
screw dislocation core structures in precipitates for very small BCC
copper-rich precipitates (1.5 nm $<$ d $<$ 3.3 nm) \cite{d-2,ther7}.
In addition to the diameter of precipitates, the temperature and
dislocation line characteristics are important to the dislocation
core-precipitate interaction \cite{ther9}. Experimental evidence
relating to the dislocation core-precipitate interaction has been
observed using transmission electron microscope (TEM) experiments.
In situ TEM demonstrates dislocations pinned and curved with obtuse
bow-out angles by BCC copper-rich precipitates with d $\sim$ 2 nm
\cite{d-3,d-4}. TEM observations also demonstrate that the
transformation of copper phase and the appearance of dislocation
loops are induced by precipitates with d $\sim$ 4 nm \cite{d-1}. The
bow-out angle of the dislocation can further be used to estimate
macroscopic hardening \cite{d-1}.

Many experiments have specifically studied the copper-rich
precipitation embrittlement effect on RPV steels and model alloys.
Positron annihilation experiments suggest strongly that copper-rich
precipitates are responsible
 for irradiation-induced embrittlement \cite{cu-brittle}. The influence of
nickel content on the embrittlement of RPV steels has also been
found to be
important\cite{cu-nin-3,cu-nin-4,cu-nin-5,cu-nin-1,cu-nin-2,cu-ni1},
and the nickel and copper content in steels have a synergistic
effect on the embrittlement tendency
\cite{miller3,miller1,cu-ni0,cu-ni0new}. It has been reported that
RPV steels with high nickel content have a higher ductile-brittle
transition temperature (DBTT) shift than low-nickel steels with same
copper content irradiated using same neutron fluence \cite{miller3}.
However, the shifts in the DBTT are small for irradiated low-copper,
high-nickel RPV steels that do not contain copper-rich precipitates
\cite{miller1}. Furthermore, a parametric study of model alloys
after neutron irradiation (at a neutron fluence of $72 \times
10^{18}m^{-2}$) showed that DBTT shifts increase with nickel content
and that the shift is visible from a threshold copper content of
approximately 0.08 at.\% \cite{cu-ni0,cu-ni0new}. These results
demonstrate clear evidence for a relation between embrittlement and
copper and nickel contents. The influences of nickel on copper-rich
precipitates have therefore attracted significant attention.

Microstructure experiments show that nickel can occur in the copper
precipitates. Atom probe experiments on thermally aged model alloys
demonstrate that the nickel is located in the core region of the BCC
copper-rich precipitates during the initial growth stage and is
rejected from the core to the interfacial region during growth and
coarsening \cite{miller4}. Observations on neutron-irradiated RPV
steels demonstrate a higher nickel concentration in the copper
precipitates than in thermally aged model alloys \cite{miller4,new}.
Recently, a series of atom probe experiments have revealed the
growing and coarsening behaviour of copper-rich precipitates in
concentrated multicomponent alloys with high strength \cite{FeCu1,
FeCu2, FeCu3, FeCu4, phd-thesis, FeCu5, FeCu6}. The nickel and other
alloying elements have been observed to segregate at the interface
of the precipitates. Moreover, phase-field \cite{FeCua1} and
Langer-Schwartz \cite{FeCua2} simulations of copper precipitates
nucleation also indicated nickel segregation at the interface of
copper precipitates during growth. First-principles calculation is
very efficient in predicting the embrittlement potential and aids in
understanding the mechanism of the impurity effect based on
electronic structure \cite{britte-bi,britte1, britte2, britte3,
britte4,review}. These findings motivated us to perform a careful,
first-principles investigation of the contribution of segregation,
especially of nickel, on the interfacial cohesive property of
precipitates to better understand the effects of these precipitates
on embrittlement.

In this paper, we study aluminium and nickel as typical alloying
elements to examine the effect of segregation at the interface of
copper-rich precipitates towards steel embrittlement. After
calculating the segregation energies, we characterise the effect of
aluminium and nickel on the interfacial cohesion of copper-rich
precipitates and attempt to explain their bonding properties in
terms of their electronic structures. Finally, we discuss the roles
of aluminium and nickel in the embrittlement of ferritic steels. Our
results suggest that the nickel-induced interfacial embrittlement
process increases the DBBT of ferritic steels.

\section{\label{sec:level1}  Methods and models}

Our first-principles calculations are based on density functional
theory (DFT) \cite{v43,v44} and performed using the Vienna Ab-initio
Simulation Package (VASP) \cite{v49} with a plane wave basis set
\cite{v45,v46}. The electron exchange and correlation is described
within the generalised gradient approximation (GGA) \cite{v47,v48},
and the interaction between ions and electrons is described using
the projector augmented wave method (PAW) \cite{v50}. The PAW
potentials we chose are treated by considering Fe3d4s, Cu3d4s,
Al3s3p and Ni3d4s as valence states. All calculations include spin
polarisation. The structural relaxations of the ions are calculated
by conjugate-gradient (CG) algorithm.

We model the coherent (001) interfaces between copper-rich
precipitates and the ferritic matrix by designing (2 $\times$ 2
$\times$ 10) multi-layered supercells composed of BCC copper and BCC
iron (Fig. 1). The lattice constants of the supercells are set using
the theoretical value for BCC iron. Our BCC iron value, 2.83 {\AA},
is reasonably consistent with the experimental value, 2.86 {\AA}
\cite{bccfe2}. These multilayer structures have a distance of 10
atomic layers between each interface, and this is determined by
considering the balance of avoidance of the interface interaction
and computational expense. The three structures illustrated in Figs.
1a-c correspond to the interfaces of precipitates with copper
concentrations of 100, 75, and 50 at.\%, respectively. The k-points
sample for these supercells are Monkhorst-Pack grids (6 $\times$ 6
$\times$ 1).

We create initial configurations of supercells modelling the
alloying element (M = Al, Ni) in different sites by substituting M
for the iron or copper atoms from (2 $\times$ 2 $\times$ 10)
supercells. We substitute M for the iron atom at site 1, to simulate
the system of M in the bulk of the ferritic matrix. The distance
from site 1 to the interface is five atomic layers, sufficient to
avoid the interaction between M and the interface. Sites 2 and 3 are
used to simulate M segregation at the interface toward the matrix
and toward the precipitated phase, respectively. Site 4 is used to
simulate M in the precipitated phase.

The formation energy is one of quantities typically used to describe
the segregation behaviour. The formation energy (E$^M$) of M in the
crystal is defined as:

\begin{equation}
  E^M = E_{M+CR} - E_{CR} - (E_{M} - E_{A})
 \end{equation}
where E$_{M+CR}$ and $E_{CR}$ are the total energies of the
supercell with a substitution defect M and defect-free supercell,
while E$_{M}$ and E$_{A}$ are the energies per atom of equilibrium
pure-element reference states. The formation energy is dependent on
the reference states, which induce arbitrariness in the result
\cite{part}. It is usually difficult to choose the correct form of
reference states when one calculates the formation energy to predict
the segregation behaviour.

Here, a more efficient quantity, the segregation energy, is used to
predict the segregation behaviour. The segregation energy
(E$^M_{X}$) of M at site X can be written as:

\begin{equation}
     \triangle E^M_{X} = E^{tot}_{M-X} - E^{tot}_{M-matrix}
\end{equation}
where E$^{tot}_{M-X}$ and E$^{tot}_{M-matrix}$ are the total
energies of the supercells for M at site X in precipitated phase and
that for M in the matrix, respectively. A negative segregation
energy indicates that M can transfer from the matrix to site X,
whereas a positive segregation energy indicates that M prefers to
dissolve within the matrix. The segregation energy reflects the
competitive capacity of trapping M between the matrix and the
precipitated phase. This strategy has previously been used to
predict partition behaviours between cementite and ferrite
\cite{part}, and the method has been verified to be appropriate.

The embrittlement property of the interface can be obtained from the
Griffith work separating the interface. Based on the Rice-Wang mode
\cite{mode}, the Griffith work is a linear function of the
difference in segregation energy for the alloying element M at the
interface ($\triangle E^M_I$) and that at the fracture free surface
($\triangle E^M_F$), $\triangle E^M_I$- $\triangle E^M_F$. An
alloying element M with positive $\triangle E^M_I$- $\triangle
E^M_F$ will reduce the cohesion of the interface and induce an
embrittlement potency, or vice versa. For the embrittlement
property, the segregation energies ($\triangle E^M_F$) at the
fracture free surface are needed. We construct an isolated fracture
free (001) ferritic matrix by subtracting the copper-rich phase from
(2 $\times$ 2 $\times$ 10) supercells representing the interface to
calculate $\triangle E^M_F$(Fig. 2).

In addition to energetic properties, we also analyse electronic
structures to provide insights into the bonding properties. We focus
on the charge density differences within the region of the interface
and the fracture free surface as shown in Fig. 2. The charge density
differences(Fig. 3) are obtained by subtracting the superimposed
charge density from the self-consistent charge density of the
relaxed structure. Fig. 3 shows the charge accumulation and
depletion, which indicate the interatomic interactions.

\section{\label{sec:level1}  Results and Discussions}

\subsection{\label{sec:level2}  Segregation energy }

Because the composition of copper-rich precipitates in ferritic
steels has a wide range, we present the segregation behaviour
properties at the interfaces of copper-rich precipitates with
different compositions. We have computed the interfaces of BCC
precipitates with 100, 75, and 50 at.\% copper. The segregation
energies of aluminium and nickel at the interfaces of precipitated
phases with different compositions are listed in Tables 1 and 2.
Apart from the total segregation energy ($\triangle E^{M}_X$), we
also present the chemical and mechanical contributions to the
segregation energy. The chemical contribution to segregation is
represented as $\triangle E^{chem,M}_X$, which is calculated based
on the unrelaxed structure. The mechanical contribution to
segregation is represented as $\triangle E^{mech,M}_X$, i.e., the
difference in segregation energy between the unrelaxed and relaxed
structures. The relaxation energies of the supercell, $\triangle
E_{sc}$, are also used when estimating the lattice distortion.

\subsubsection{\label{sec:level1} Aluminium segregation}

The segregation energies (see $\triangle E^{Al}_X$ in Table 1) of
aluminium at the interfaces of the precipitated phases are predicted
to be negative, indicating that the presence of aluminium atoms at
the interfaces is energetically favourable compared to that in the
ferritic matrix. Moreover, the segregation energies of aluminium at
the interfaces are lower than those in the core regions of the
precipitates, indicating that the presence of aluminium atoms at the
interface is more favourable than that in the core region. These
results indicate that the interface between the matrix and the
precipitated phase can trap aluminium atoms, consistent with
three-dimensional atom probe (3DAP) experiments \cite{FeCu1, FeCu2,
FeCu3, phd-thesis}.

The interfacial segregation energy at the interface for precipitates
with 100 at.\% copper is larger than that for the precipitated phase
with 50 at.\% copper by 1.56 eV. This result indicates that
interfacial segregation is strongly dependent on the composition of
the precipitated phase and increases with its copper concentration.
The strong dependence of interfacial segregation on the composition
of precipitates plays an important role in experimental phenomena in
which the concentration of aluminium at the interface of
precipitates increases in ferritic steels during thermal treatment
processes \cite{FeCu1,FeCu2,phd-thesis}.

\subsubsection{\label{sec:level1}  Nickel segregation  }

We find that the nickel segregation behaviour is similar to that of
aluminium. The segregation energies (see $\triangle E^{Ni}_X$ in
Table 2) of nickel in the interfaces are negative, indicating that
the presence of nickel atoms in the interfaces is energetically
favourable compared to that in the ferritic matrix. However, the
most favourable sites depend on the composition of the precipitated
phase. The most favourable sites are in the core region for the
precipitated phase with 50 at.\% copper, but at the interface for
the precipitated phase with 100 and 75 at.\% copper. These results
indicate that nickel atoms can partition at the precipitates with 50
at.\% copper and segregate into the interfaces of the precipitated
phase with 100 and 75 at.\% copper in a ferritic matrix. Therefore,
nickel will segregate into the core region of the precipitates at
the initial formation stage and be pushed away from the core towards
the interface of the precipitates for the following growth stage.
This is consistent with phase field \cite{FeCua1} and
Langer-Schwartz \cite{FeCua2} simulations. These phenomena were also
observed for the copper precipitates in RPV steels using 3DAP
\cite{miller4,new}.

The segregation energy of nickel at the interface also depends on
the composition of the precipitated phase. The segregation energy at
the interface of the precipitated phase with 100 at.\% copper is
larger than that of 50 at.\% copper by 0.22 eV, a difference much
lower than that of aluminium. This result indicates that the nickel
concentration at the interface of copper-rich precipitates increases
more slowly than that of aluminium with increasing copper
concentration during the growth of precipitates. This trend has
previously been observed in 3DAP experiments \cite{concentration}.

\subsubsection{\label{sec:level1}  Chemical and mechanic
contributions }

As shown in Tables 1 and 2, the values of  $\triangle E^{chem}_X$
approach those of $\triangle E_X$, so long as the $\triangle
E^{mech}_X$  is very small. These values indicate that chemical
energy is the main contributor to the segregation behaviour of
aluminium and nickel. $\triangle E_{sc}$  and $\triangle E_X$  are
of comparable magnitude, implying that the relaxation energy per
atom is much smaller than the interfacial segregation energy. The
average relaxation energies per copper atom for precipitated phases
with 100, 75 and 50 at.\% copper are 0.012 eV, 0.027 eV, and 0.013
eV, whereas the smallest segregation energies of aluminium and
nickel are 0.15 eV and 0.20 eV, respectively. These values reflect
the fact that the energy gains of the interfacial segregation of
aluminium and nickel are much larger than that of lattice
distortion.

It is necessary to discuss the factors affecting the relaxation
energy of lattice distortion. The relaxation energy is mainly
determined by mechanical stability and the mixing effects of the BCC
FeCu metastable alloy. The mechanical stability and mixing effect
both decrease with copper concentration (at copper concentrations
$>$ 50 at.\%) \cite{fecuphase,xie-zhao}. Higher mechanical stability
reduces the relaxation energy, whereas larger mixing effects enhance
the relaxation energy. Therefore, the relaxation energy of the
precipitated phase with 75 at.\% copper become the largest among the
three structures studied due to the compromise formed between the
effects of mechanical stability and mixing.

\subsection{\label{sec:level1} The Griffith work is influenced by segregation}

We take the interface between the pure copper and ferritic matrix as
a typical mode to analyse the Griffith work. The Griffith work is
the energy separating an interface against the atomic cohesion. The
influence of alloying element segregation on the Griffith work can
be estimated by the value of $\triangle E^M_I$- $\triangle E^M_F$.
We have calculated the segregation energies ($\triangle E^M_I$) at
the interface of precipitates in ferritic steels previously. Now, we
calculate the segregation energies ($\triangle E^M_F$) of aluminium
and nickel at the fracture free surfaces, and the results are -0.78
and -0.51 eV, respectively. The value of $\triangle E^M_I$-
$\triangle E^M_F$ for aluminium and nickel at the interface between
pure copper and the ferritic matrix are -0.93 eV and 0.07 eV,
respectively. As a result, the segregation of aluminium will enhance
the interfacial cohesion, whereas the segregation of nickel will
reduce the interfacial cohesion.

First, we discuss the electronic structures of aluminium and nickel
segregation at the interface. The charge differences of aluminium
and nickel at the interface are presented in the left-hand column of
Fig. 3, showing that the charge accumulates in the interval region
between atoms and is depleted in the inner atomic shells. The charge
depletion region for aluminium atoms is larger than that of iron and
copper atoms, because the 3s3p electrons of aluminium are more
delocalised than the 3s3d electrons of iron and copper atoms. The 3p
electrons of aluminium fill into the degeneration state of p$_x$,
p$_y$, p$_z$, therefore the charge depletion region for the
aluminium atom has a higher symmetry pattern. In contrast, the
charge depletion region for the nickel atom is similar to that of
iron and copper atoms, because the valence states of nickel also
include 3s3d. Because the delocalised electrons can be affected more
by surrounding atoms, the segregation energy of aluminium is larger
than that of nickel and more sensitive to the composition of the
precipitated phase.

We now attempt to understand the effect of segregation on Griffith
work by comparing the chemical bonding in the fracture free surface
to that in the interface. The charge differences of aluminium and
nickel at the fracture free surface are presented in the right-hand
column of Fig. 3. Because the geometrical symmetry is broken for the
free surface, the orbital p$_z$ of aluminium state hybridises with
the d$_{z^2}$ of iron and becomes lower in energy. The only one
p-electron of aluminium fills into p$_z$, and the p$_{xy}$ is left
unoccupied. This results stronger vertical bonding (Fe2-Al) and
weaker lateral chemical bonding (Fe1-Al and Fe3-Al) of the free
surface compared to that of the interface. The weakening effect of
aluminium on lateral bonding contributes to a lower segregation
energy at the free surface than that at the interface (by 0.93 eV).
Therefore, $\triangle E^{Al}_I$- $\triangle E^{Al}_F$ for the
segregation of aluminium is negative.

The alteration of the chemical bonding of nickel is totally
different to that of aluminium. The spatial distribution of nickel
at the fracture free surface is similar to that at the interface due
to the d electron. The charge accumulations in the interval region
of Fe1-Ni, Fe2-Ni, and Fe3-Ni for the free surface are all greater
than that for the interface, indicating a stronger chemical bonding
between these atoms at free surface. The enhanced chemical bonding
arises from the contraction of bond lengths at the free surface.
Because the chemical bonding at the free surface is stronger than at
the interface, the segregation energy at the free surface is larger
than at the interface (by 0.07 eV). $\triangle E^{Ni}_I$- $\triangle
E^{Ni}_F$ for the segregation of nickel is consequently positive.

\subsection{\label{sec:level1}  The embrittlement trend}

The influence of alloying element M on an interfacial cohesive
property is determined by both $\triangle E^M_I$ and $\triangle
E^M_I$- $\triangle E^M_F$. We plot $\triangle E^M_I$ and $\triangle
E^M_I$- $\triangle E^M_F$ for precipitated phases with different
compositions in Fig. 4, showing that the values of $\triangle
E^{Al}_I$ and $\triangle E^{Al}_I$- $\triangle E^{Al}_F$ strongly
depend on the composition of the precipitated phase. The values of
$\triangle E^{Al}_I$- $\triangle E^{Al}_F$ are positive for
precipitated phases of 50 $\sim$ 65 at.\% copper and negative for
precipitated phases of 65 $\sim$ 100 at.\% copper. Obviously, there
is a wide range of compositions for the precipitates whose
interfacial cohesion are enhanced by aluminium segregation. The
values of $\triangle E^{Al}_I$ for these precipitates are all larger
than those of the precipitates whose interfacial cohesion is reduced
by aluminium segregation. Therefore, aluminium segregation plays a
prominent role in enhancing interfacial cohesion for copper-rich
precipitates in ferritic steels.

In contrast, the values of $\triangle E^{Ni}_I$  and $\triangle
E^{Ni}_I$- $\triangle E^{Ni}_F$ weakly depend on the composition of
the precipitated phase. The value of $\triangle E^{Ni}_I$  varies
weakly with the composition of the precipitated phase. The values of
$\triangle E^{Ni}_I$- $\triangle E^{Ni}_F$ are all negative. These
results indicate that nickel segregation can reduce the interfacial
cohesion of copper-rich precipitated phases of any composition.
Undoubtedly, nickel segregation plays a constant role in reducing
interfacial cohesion for copper-rich precipitates in ferritic
steels.

BCC copper-rich precipitates play important roles in dislocation
pining and misfit growth in ferritic steels. Dislocation pinning
will strengthen the ferritic matrix, whereas the misfit growth will
improve the ductility. The influence of the misfits produced by
copper precipitates on ductile-brittle transformation has been
proven to be considerable \cite{fine1,fine2}. The interfacial
cohesive property and the amounts of precipitates present contribute
importantly to ductile-brittle transformation. The ductile-brittle
transition depend the competition between fracture stress and flow
stress. Flow stress increases with decreasing temperature. When the
flow stress is larger than the fracture stress at lower
temperatures, the ferritic matrix is brittle; and when the flow
stress is smaller than the fracture stress at higher temperatures,
the ferritic matrix is ductile. Therefore, the DBTT can be altered
by fracture stress, which is determined by the interfacial cohesion
of precipitates. We can now predict that the segregation of
aluminium can lower the DBTT due to the enhancing fracture stress of
copper precipitates, whereas the segregation of nickel can increase
the DBTT due to the reducing fracture stress of copper precipitates.

The nickel-induced interfacial embrittlement of copper-rich
precipitates explains the observation that the DBTT of low-carbon,
copper-precipitation-strengthened steels increase with nickel and
copper content \cite{fine1}. This effect also accounts for the
observation that the shifts of DBTT in RPV steels after neutron
irradiation are enhanced by the copper and nickel content
\cite{miller3,miller1}. Furthermore, it can account for the
observations that the influence of copper content on DBTT decrease
is progressive when the nickel content decreases and the influence
of nickel content on DBTT disappears for model alloys after neutron
irradiation (at a neutron fluence of $72 \times 10^{18}m^{-2}$) with
copper contents below 0.08 at.\% (0.1 wt.\%)
\cite{cu-ni0,cu-ni0new}.

\section{\label{sec:level1}  Conclusion}

We have presented DFT-GGA calculations investigating the segregation
behaviours of aluminium and nickel at the interface of precipitates
in ferritic steels. We have examined the segregation energies of
aluminium and nickel at the interface and within the core regions in
precipitated phases with different compositions. Our results show
that aluminium and nickel can segregate at the interface of
precipitates in ferritic steels, in agreement with 3DAP experiments.
Moreover, we also find that the interfacial segregation of aluminium
is more sensitive to the composition of copper-rich precipitated
phases than those of nickel. The most energetically favourable site
of nickel segregation depends to the composition of the copper-rich
precipitated phase. These detailed segregation behaviours are also
consistent with 3DAP experiments.

We also calculated the contributions of aluminium and nickel
segregation behaviours to Griffith work to predict the interfacial
cohesive properties of precipitates and found that there are
aluminium-induced ductility and nickel-induced embrittlement effects
at the interface of precipitates. Aluminium-induced ductility arises
mainly from the 3p electron of aluminium that causes weaker Fe-Al
bonding at the fracture surface than at the interface.
Nickel-induced embrittlement, however, is due mainly to the 3d
electrons of nickel resulting in enhanced Fe-Ni bonding at the
fracture surface compared to that at the interface. Finally, we have
used the nickel-induced embrittlement at the interface of the copper
precipitates in ferritic matrices to explain the experimental
observation that the DBTT of low-carbon,
copper-precipitation-strengthened steels and RPV steels increase
with their nickel and copper content. These studies suggest a
possibility of improving the ductility of ferritic steels from
modifying interfacial cohesive properties of copper-rich
precipitates by segregation of alloying elements.

The authors thank Professor B. X. Zhou, Mr G. Xu and Q. D. Liu for
critical discussions. This work is financially supported by National
Science Foundation of China (Grant No. 50931003, 51001067), Shanghai
Committee of Science and Technology (Grant No. 09520500100),
 Shanghai Municipal Education Commission (Shu-Guang project, Grant No. 09SG36) and Shanghai
Education Development Foundation and Shanghai Leading Academic
Discipline Project (S30107). The computations were performed at
Ziqiang Supercomputer Center of Shanghai University and Shanghai
Supercomputer Center.

\begin{table}[b]
\caption{\label{tab:fonts} Segregation energies (in eV) of Al atoms
at the interfaces and core regions of Cu-rich precipitated phases in
a ferritic matrix, decompositions into chemical and mechanical
contributions, and the relaxation energy of the supercell.
 }

\begin{ruledtabular}
\begin{tabular}{c@{~~~~~~}c@{~~~~}c@{~~~~~~}c@{~~~}c}

Site(X) &  1 &  2 & 3 & 4 \\[0.5ex]
 \hline

\multicolumn{5}{c} {\itshape $\alpha$-Fe/prue BCC-Cu }\\[0.5ex]
$\triangle E^{chem,Al}_X$& 0  & -1.78  & -1.52  & -1.18 \\[0.5ex]
$\triangle E^{mech,Al}_X$&   0 &  +0.07& +0.07 &  +0.06  \\[0.5ex]
$\triangle E^{Al}_X$&   0 &  -1.71& -1.45 &  -1.12  \\[0.5ex]

$\triangle E_{sc}$ & -0.60   & -0.53  & -0.53  & -0.54 \\[0.5ex]

\multicolumn{5}{c} {\itshape $\alpha$-Fe/BCC-CuFe( 75 at.\%  Cu) }\\[0.5ex]
$\triangle E^{chem,Al}_X$& 0 &  -1.07 &  -0.93 &   -0.60\\[0.5ex]
$\triangle E^{mech,Al}_X$& 0 &  -0.01  &  +0.02  &-0.07 \\[0.5ex]
$\triangle E^{Al}_X$&   0  & -1.08& -0.91&   -0.67  \\[0.5ex]
$\triangle E_{sc}$ & -0.84 &  -0.85  & -0.82   &-0.91 \\[0.5ex]

\multicolumn{5}{c} {\itshape $\alpha$-Fe/BCC-CuFe( 50 at.\%  Cu) }\\[0.5ex]
$\triangle E^{chem,Al}_X$& 0  & -0.12  & -0.13   &0.14 \\[0.5ex]
$\triangle E^{mech,Al}_X$&  0 &  -0.03  & -0.03  & -0.01 \\[0.5ex]
$\triangle E^{Al}_X$&  0& -0.15& -0.16 &  0.13 \\[0.5ex]
$\triangle E_{sc}$&  -0.28 &  -0.31  & -0.31  & -0.29 \\[0.5ex]

\end{tabular}
\end{ruledtabular}
\end{table}

\begin{table}[b]
\caption{\label{tab:fonts} Segregation energies (in eV) of Ni atoms
at the interfaces and core regions of Cu-rich precipitated phases in
a ferritic matrix, decompositions into chemical and mechanical
contributions, and the relaxation energy of the supercell. }
\begin{ruledtabular}
\begin{tabular}{c@{~~~~~~}c@{~~~~}c@{~~~~~~}c@{~~~}c}

Site(X) &  1 &  2 & 3 & 4 \\[0.5ex]
\hline
\multicolumn{5}{c} {\itshape  $\alpha$-Fe/pure BCC-Cu    }\\[0.5ex]
$\triangle E^{chem,Ni}_X$& 0  &  -0.39 &  -0.42  & -0.05 \\[0.5ex]
$\triangle E^{mech,Ni}_X$& 0  & -0.05 &  -0.01  & -0.02 \\[0.5ex]
$\triangle E^{Ni}_X$&  0  & -0.44 & -0.43& -0.03 \\[0.5ex]
$\triangle E_{sc}$& -0.43& -0.48 &  -0.44  & -0.41 \\[0.5ex]

\multicolumn{5}{c} {\itshape $\alpha$-Fe/BCC-CuFe( 75 at.\%  Cu) }\\[0.5ex]
$\triangle E^{chem,Ni}_X$& 0 &  -0.37 &  -0.33  & -0.15 \\[0.5ex]
$\triangle E^{mech,Ni}_X$& 0  & -0.03  & -0.04  & -0.03 \\[0.5ex]
$\triangle E^{Ni}_X$&   0  & -0.40& -0.38 &  -0.18 \\[0.5ex]
$\triangle E_{sc}$& -0.73  & -0.76  & -0.78  & -0.76 \\[0.5ex]

\multicolumn{5}{c} {\itshape $\alpha$-Fe/BCC-CuFe( 50 at.\%  Cu) }\\[0.5ex]
$\triangle E^{chem,Ni}_X$& 0  & -0.22 &  -0.24 &  -0.29 \\[0.5ex]
$\triangle E^{mech,Ni}_X$&  0  & -0.02  & -0.03  & +0.01  \\[0.5ex]
$\triangle E^{Ni}_X$&  0  & -0.20& -0.27 &-0.28 \\[0.5ex]
$\triangle E_{sc}$&  -0.24  & -0.22  & -0.27  & -0.23  \\[0.5ex]

\end{tabular}
\end{ruledtabular}
\end{table}

\begin{figure*}[b]
\includegraphics[scale=1.6,angle=0]{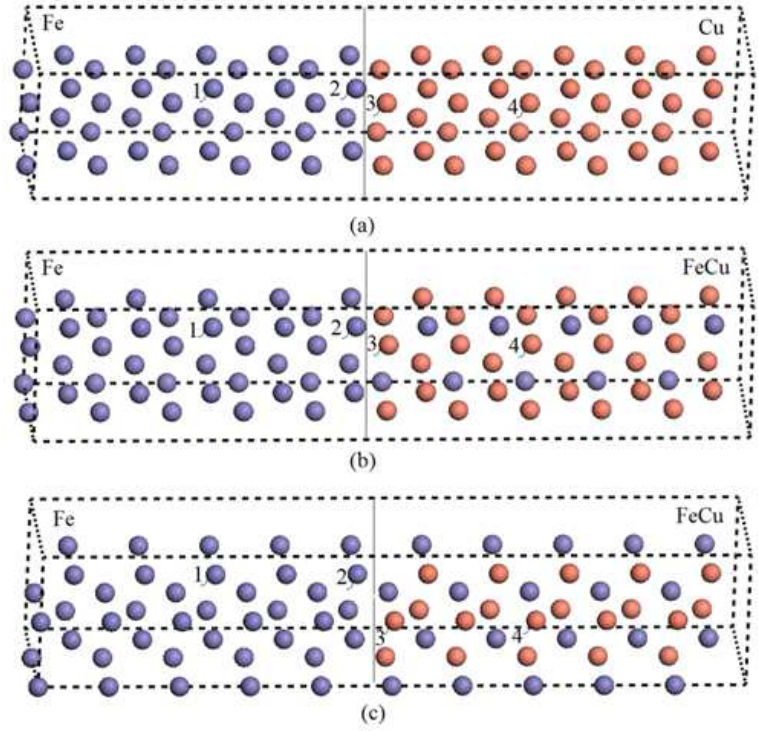}
\caption{ The atomic structures of the (001) interfaces between BCC
Cu-rich precipitated phases and ferritic matrix. The precipitated
phases are Cu-Fe alloys with (a) 100 at.\% (i.e., pure Cu), (b) 75
at.\%, and (c) 50 at.\% Cu concentration, respectively. Red and blue
balls denote Cu and Fe atoms, respectively. The atoms marked by
Arabic numerals denote the segregation sites in 1, the matrix; 2,
the region of interface toward the matrix; 3, the region of
interface toward the precipitated phase; and 4, the core region of
the precipitated phase.
 }\label{fig1}
\end{figure*}

\begin{figure*}[b]
\includegraphics[scale=0.6,angle=0]{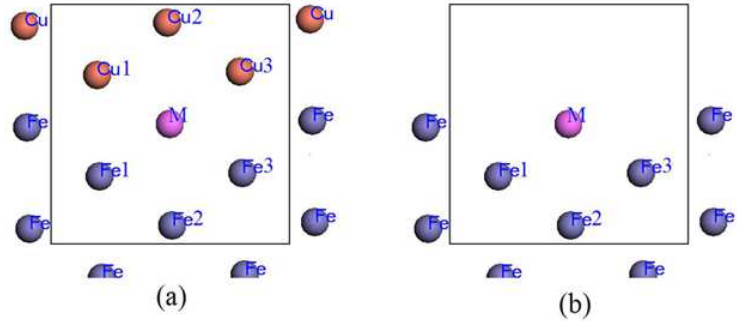}
\caption{ Side view of the atomic structures of M segregation at (a)
the interface between the BCC Cu-rich precipitated phase and the
ferritic matrix and (b) its corresponding fracture free surface. }
\label{fig1}
\end{figure*}

\begin{figure*}
\includegraphics[scale=0.82,angle=0]{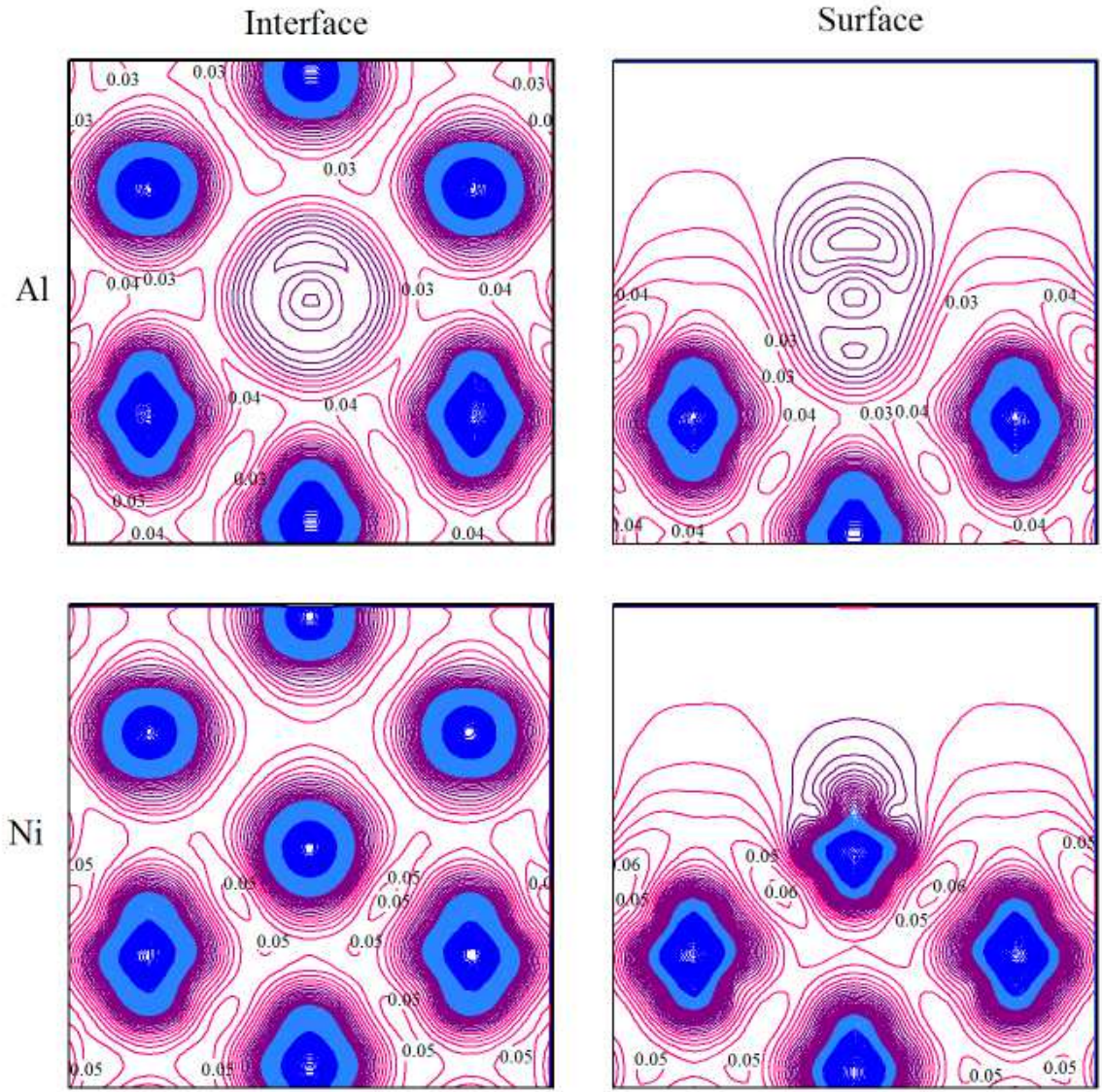}
\caption{ The valence charge density differences of Ni at the
interface (top left), Ni at the fracture free surface (top right),
Al at the interface (bottom left), and Al at the fracture free
surface (bottom right). Contours increase successively by a factor
of 10$^{-2}$/au$^3$. Blue, light blue, and purple lines denote
charge depletion and pink lines denote charge accumulation. Regions
displayed correspond to the square cell depicted in Fig. 2.
}\label{fig1}
\end{figure*}

\begin{figure*}[b]
\includegraphics[scale=0.30,angle=0]{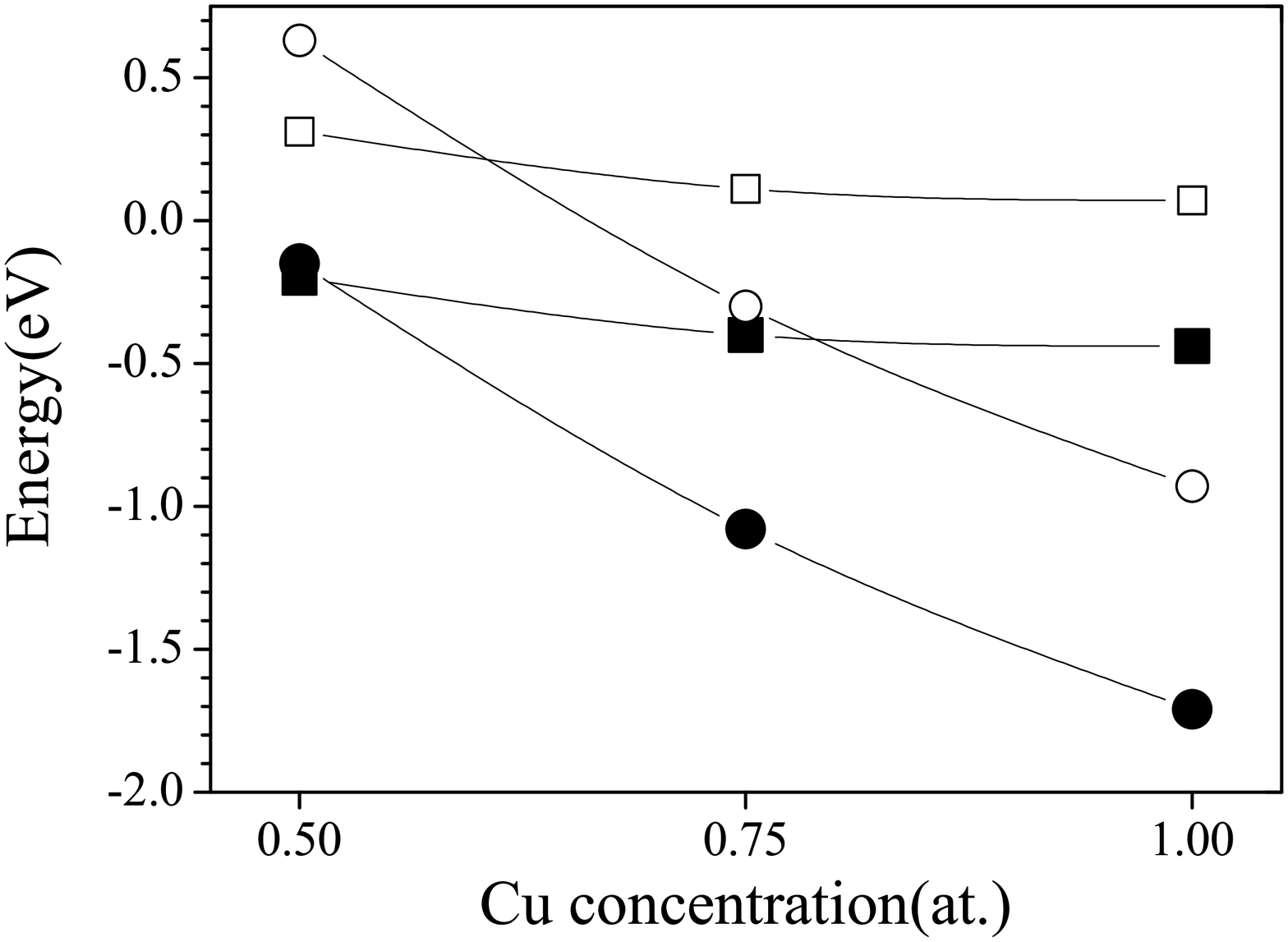}
\caption{
 $\triangle E^M_I$ (solid symbols) and  $\triangle
E^M_I-\triangle E^M_F$ (blank symbols) for Ni (squares) and Al
(dots) segregating at the interfaces of the copper-rich precipitated
phases with different copper concentrations in the ferritic matrix.
} \label{fig1}
\end{figure*}



\begin{thebibliography}{}

\bibitem{steel-high1} S. K. Dhua, D. Mukerjee, and D. S. Sarma, Metall. Mater. Trans. A, 32, 2259 (2001).
\bibitem{steel-high2} S. Vaynman, D. Isheim, R. P. Kolli, S. P. Bhat, D. N. Seidman, and M. E. Fine, Metall. Mater. Trans. A, 39, 363 (2008).
\bibitem{steel-high3} X. Yu, J. L. Caron, S. S. Babu, J. C. Lippold, D. Isheim, and D. N. Seidman, Acta. Mater., 58, 5596 (2010).
\bibitem{miller2} W. J. Phythian, and C. A. English, J. Nucl. Mater., 205, 162 (1993).
\bibitem{rpv2} E. A. Kuleshova, B. A. Gurovich, Y. I. Shtrombakh, Y. A. Nikolaev, and V. A. Pechenkin, J. Nucl. Mater., 342, 77 (2005).
\bibitem{rpv3} M. K. Miller, and K. F. Russell, J. Nucl. Mater., 371, 145 (2007).
\bibitem{ther2} T. Harry and D. J. Bacon,  Acta. Mater., 50, 195 (2002).
\bibitem{ther3} T. Harry and D. J. Bacon,  Acta. Mater., 50, 209 (2002).
\bibitem{conment} M. E. Fine and D. Isheim, Scripta Mater 53, 115 (2005).
\bibitem{d-2} J. Shim, Y. Cho, S. Kwon, W. Kim, and B. Wirth, Appl. Phys. Lett., 90, 021906 (2007).
\bibitem{ther7} Z. Z. Chen, N. Kioussis, and N. Ghoniem, Phys. Rev. B 80, 184104 (2009).
\bibitem{ther9} G. Monnet, S. Naamane, and B. Devincre, Acta. Mater., 59, 451 (2011).
\bibitem{d-3} K. Nogiwa, T. Yamamoto, K. Fukumoto, H. Matsui, Y. Nagai, K. Yubuta, and M. Hasegawa, J. Nucl. Mater., 307, 946 (2002).
\bibitem{d-4} K. Nogiwa, N. Nita, and H. Matsui, J. Nucl. Mater., 367, 392 (2007).
\bibitem{d-1} S. Lozano-Perez, G. Sha, J. M. Titchmarsh, M. L. Jenkins, S. Hirosawa, A. Cerezo, and G. D. Smith, J. Mater. Sci., 41, 2559 (2006).
\bibitem{cu-brittle} Y. Nagai, Z. Tang, M. Hassegawa, T. Kanai, and M. Saneyasu, Phys. Rev. B., 63, 134110 (2001).
\bibitem{cu-nin-3} M. M. Ghoneim and F. H. Hammad, Int. J. Pres. Ves and Piping, 74, 189 (1997).
\bibitem{cu-nin-4} A. M. Kryukov and Y. A. Nikolaev,  Nucl. Eng. Des., 195, 143 (2000).
\bibitem{cu-nin-5} R. Ahlstrand, M. Bi$\grave{e}$th, and C. Rieg, Nucl. Eng. Des., 230, 267 (2004).
\bibitem{cu-nin-1} P. Efsing, C. Jansson, T. Mager and G. Embring, ASTM international, 4, 1 (2007).
\bibitem{cu-nin-2} Y. A. Nikolaev,  ASTM international, 4, 56 (2007).
\bibitem{cu-ni1} W. D. Yang, Reactor Material Science (in chinese), (Atomic Energy Press,
2000).
\bibitem{miller3} M. K. Miller, K. F. Russell, M. A. Sokolov, and R. K. Nanstad, J. Nucl. Mater. 361, 248 (2007).
\bibitem{miller1} M. K. Miller, M. A. Sokolov, R. K. Nanstad, and K. F. Russell, J. Nucl. Mater. 351, 187 (2006).
\bibitem{cu-ni0} L. Debarberis, F. Sevini, B. Acosta, A. Kryukov,
Y. Nikolaev, A. D. Amaev, and M. Valo, Int. J. Pressure Vessels
Piping 79, 637 (2002).
\bibitem{cu-ni0new} B. Acosta, L. Debarberis, F. Sevini, and A. Kryukov, NDT. E. International 37, 321 (2004).
\bibitem{miller4} P. J. Pareige, K. F. Russell, and M. K. Miller, Appl. Surf.
Sci. 94, 362 (1996).
\bibitem{new} J. M. Hyde, G. Sha, E. A. Marquis, A. Morley, and K. B. Wilford, J. J. Williams, Ultramicroscopy,
(doi:10.1016/j.ultramic.2010.12.030)
\bibitem{FeCu1} D. Isheim, M. S. Gagliano, M. E. Fine, and D. N. Seidman, Acta. Mater. 54, 841 (2006).
\bibitem{FeCu2} D. Isheim, R. P. Kolli, M. E. Fine, and D. N. Seidman, Scripta Mater 55, 35 (2006).
\bibitem{FeCu3} R. P. Kolli, and D. N. Seidman, Microsc Microanal 13, 272 (2007).
\bibitem{FeCu4} R. Kolli, Z. Mao, D. N. Seidman, and D. T. Keane, Appl. Phys. Lett. 91, 241903 (2007).
\bibitem{phd-thesis} R. P. Kolli, ¡°Kinetics of nanoscale Cu-rich precipitates in a
multicomponent concentrated steel,¡± (Ph.D. thesis, Northwestern
University, 2007).
\bibitem{FeCu5} R. P. Kolli, and D. N. Seidman, Acta. Mater. 56, 2073 (2008).
\bibitem{FeCu6} R. P. Kolli, R. M. Wojes, S.  Zaucha, and D. N. Seidman, Int. J. Mater. Res. 99, 513 (2008).
\bibitem{FeCua1} T. Koyama, K. Hashimoto, and H. Onodera, Mater. Trans., 47, 2765 (2006).
\bibitem{FeCua2} C. Zhang and M. Enomoto, Acta. Mater. 54, 4183 (2006).
\bibitem{britte-bi} R. W. Wu, A. J. Freeman, and G. B. Olson, Science, 265, 376 (1994).
\bibitem{britte1} Y. Liu, K. Y. Chen, G. Lu, J. H. Zhang, and Z. Q. Hu, Acta. Mater. 45, 1837 (1997).
\bibitem{britte2} P. Peng, D. W. Zhou, J. S. Liu, R. Yang, and Z. Q. Hu, Mat. Sci. Eng. a-Struct. 416, 169 (2006).
\bibitem{britte3} Y. X. Wu, X. Y. Li, and Y. M. Wang, Acta. Mater. 55, 4845 (2007).
\bibitem{britte4} C. Wang and C. Y. Wang, Surf. Sci. 602, 2604 (2008).
\bibitem{review}  Y. Mishin, M. Asta, and J. Li, Acta. Mater. 58, 1117 (2010).
\bibitem{v43} P. Hohenberg and W. Kohn, Phys. Rev. 136, B864 (1964). W. Kohn and L. J. Sham, Phys. Rev. 140, A1133 (1965).
\bibitem{v44} R. O. Jones and O. Gunnarsson, Rev. Mod. Phys. 61, 689 (1989).
\bibitem{v49} G. Kresse and J. Furthm\"{u}ller, Phys. Rev. B, 54, 11169 (1996). Comp. Mater. Sci. 6, 15 (1996).
\bibitem{v45} R. Car and M. Parrinello, Phys. Rev. Lett. 55, 2471 (1985).
\bibitem{v46} M. C. Payne, M. P. Teter, D. C. Allan, T. A. Arias, and J. D. Joannopoulos, Rev. Mod. Phys. 64, 1045 (1992).
\bibitem{v47} Y. Wang and J. P. Perdew, Phys. Rev. B, 44, 13298 (1991).
\bibitem{v48} J. P. Perdew, J. A. Chevary, S. H. Vosko, K. A. Jackson, M. R. Pederson, D. J. Singh, and C. Fiolhais, Phys. Rev. B 46, 6671 (1992).
\bibitem{v50} G. Kresse and D. Joubert, Phys. Rev. B 59, 1758 (1999).
\bibitem{bccfe2} C. Kittel, Introduction to solid state physics, 7th
ed, (New York, Wiley, 1996).
\bibitem{part}  C. K. Ande  and  M. H. F. Sluiter, Acta. Mater., 58, 6276 (2010).
\bibitem{mode}  R. Rice and J. S. Wang , Mat. Sci. Eng. a-Struct., 107, 23 (1989).
\bibitem{concentration}  R. P. Kolli, ¡°Kinetics of nanoscale Cu-rich precipitates in a
multicomponent concentrated steel,¡± p152, 184, (Ph.D. thesis,
Northwestern University, 2007).
\bibitem{fecuphase} J. Z. Liu, A. van de Walle, G. Ghosh, and M. Asta, Phys. Rev. B, 72, 144109 (2005).
\bibitem{xie-zhao} Y. P. Xie and S. J. Zhao, to be published
\bibitem{fine1} M. E. Fine, S. Vaynman, D. Isheim, Y. W. Chung, S. P. Bhat, and C. H. Hahin, Metall. Mater. Trans. A, 41, 3318 (2010).
\bibitem{fine2} W. J. Lee, W. J. Chia, J. L. Wang, Y. F. Chen, S. Vaynman, M. E. Fine, and Y. W. Chung, Langmuir 26, 16254 (2010).





\end{thebibliography}
\end{document}